# The Dependent Chip Model (DCM): a simple and more realistic alternative to the Independent Chip Model (ICM)

E. Besalú

Department of Chemistry and

Institute of Computational Chemistry and Catalysis

University of Girona

Catalonia, Spain

DependentChipModel@gmail.com

**Abstract**

The Dependent Chip Model (DCM) is proposed as an alternative to the Independent Chip Model (ICM) usually employed in poker tournament negotiations. DCM constitutes a recursive exploration of a multiplayer Texas hold'em poker game tree tracking. The DCM procedure considers all players as having exactly the same playing skills and probabilities to win a single poker hand, but submitted to their stacks in order to survive along the successive hands. So, the final differences among bestowed prizes arise purely from the initial chip amounts, hence the name of this new proposed procedure. By considering DCM, the first podium positions usually collect more money than the amount granted by ICM. Conversely, the last podium positions receive less money than the quantities proposed by ICM. The differences among both methods sometimes lead to take distinct actions in tournaments. This can lead to very important monetary implications for professional players.

## 1. Introduction

In Texas hold'em poker tournaments it is common to have to decide in a negotiation framework for the final prizes sharing. In general, the goal is to put into correspondence the amounts of chips with money. The most known and considered procedure relies on the Independent Chip Model (ICM), a popular scheme, originated in 1986, as cited by Diaconis and Ethier (2020), and Gilbert (2009). The ICM procedure consists in generating all the possible final podium results and, for each option, calculate the probabilities of the sequence of players within an independent probabilistic framework where the probabilities of each event (i.e., to be at a certain podium position given there are other particular players at the previous and better podium steps) are calculated from the actual chips of the remaining players, following simple Laplace quotients (i.e. probability terms attached to independent events).

Diaconis and Ethier remark a desired characteristic of ICM: In a fair coin-tossing play involving two players having stacks of $A$ and $B$ chips, the derived chances of winning according to the ICM are $A/(A+B)$ and $B/(A+B)$, respectively. Miltersen and Sørensen (2007), and Ganzfried and Sandholm (2008) comment that the real probabilities approach to these quotients, as pointed by Sklansky (2002). At this point it can be notified in advance that the Dependent Chip Model (DCM) which is promoted here also fulfills this condition. So, when comparing both methods money and chips directly correlate when two players are confronted. Differences arise between ICM and DCM protocols when three or more players dispute final prizes.

As it is known for the ICM, and it will be also shown here for the DCM one, none of both methods take into account many nuances of poker (blind positions, kind of players and playing styles, effect of decreasing number of players at the table, behavior at the bubble, final strategic actions in front of other player all-ins, table dynamics, ...).

## 2. The ICM procedure

In this section the ICM procedure is described and some simple examples are given for the unique cases where ICM and DCM coincide in results: when only two players are confronted or when only a single first prize is offered.

If $N$ players remain in play with stacks $S_1$, $S_2$, ..., $S_N$; and $Z \leq N$ prizes ($C_1 \geq C_2 \geq ... \geq C_Z$) are to be shared, the ICM procedure calculates the expected payoff as follows:

- Generate the $V(N,Z) = N(N-1)(N-2)\cdots(N-Z+1)$ variations without repetition which describe all the possible podiums (ties at podium positions are not considered).
- For each variation, say $w(1)w(2)...w(Z)$ where $w(i)$ stands for the identity of the player at podium positon number $i$ do
    - The expected return to player $w(1)$ is $C_1 p(1)$, where $p(1)=S_{w(1)}/S$ is the (assumed) expected probability that player $w(1)$ finishes at position 1, and $S = \sum_{i=1}^{N} S_i$ is the sum of all the stacks.
    - The expected return to player $w(2)$, provided that player $w(1)$ already has his/her prize, is $C_2 p(1)p(2)$, where $p(2)=S_{w(2)}/(S-S_{w(1)})$. Note how this last probability is calculated only considering the stacks of player $w(2)$ and the following ones, ignoring player $w(1)$.
    - The expected return to player $w(3)$, provided that players $w(1)$ and $w(2)$ got their prizes, is $C_3 p(1)p(2)p(3)$ where $p(3)=S_{w(3)}/(S-S_{w(1)}-S_{w(2)})$, now ignoring the stacks of previous players $w(1)$ and $w(2)$.
    - ...
    - The expected return to player $w(i)$, provided that previous players $w(1)$, $w(2)$, ..., $w(i-1)$ already got their prizes, is $C_i \prod_{j=1}^{i} p(j)$, where

$$p(i) = \frac{S_{w(i)}}{S - \sum_{j=1}^{i-1} S_{w(j)}} = \frac{S_{w(i)}}{\sum_{j=i}^{N} S_{w(j)}}.$$

- And the process continues for a variation while there are players and non-null prizes to distribute.

All the expected returns are cumulated for each player when revisiting all the generated variations. The final payoffs are obtained at the end of the procedure, when all the variations are inspected and the contributions added.

| Players | Chip amounts and percentage of prize returns | | | | |
| --- | --- | --- | --- | --- | --- |
| | Player 1 | Player 2 | Player 3 | Player 4 | Player 5 |
| 2 | 1000 | 500 | | | |
| | *66.67* | *33.33* | | | |
| 3 | 5000 | 2000 | 500 | | |
| | *66.67* | *26.67* | *6.67* | | |
| 4 | 3500 | 1200 | 700 | 100 | |
| | *63.64* | *21.82* | *12.73* | *1.82* | |
| 5 | 5000 | 4000 | 3000 | 2000 | 1000 |
| | *33.33* | *26.67* | *20.00* | *13.33* | *6.67* |

**Table 1.** Coincident results for ICM and DCM (case of a single first prize). The first row entry is the number of chips of each player. The first and unique prize is of 100 monetary units. The second row in italics gives the expected money returns.

If there is only a single first prize, it is immediate to see that the ICM expected return for each player is equal to its amount of chips divided by the sum of all the stacks. For the case of two players (players 1 and 2 having $A$ and $B$ chips, respectively) there are only two possible podiums: Player 1 – Player 2 or Player 2 – Player 1. If there is only a single first prize ($Z=1$), the relevant place is the first one and, in accordance with the ICM procedure, the $p(1)$ probabilities to reach the first podium position are $A/(A+B)$ and $B/(A+B)$, respectively. The expected money return to each player are these probabilities times the amount of the first prize. The same is extensible to the many players case (when $Z=1$). This is so because, when revisiting all the podium variations, each player is placed the same number of times at the first podium position. The probability to reach such a position is considered to be, by construction, its amount of chips divided by the sum of chips of all the stacks. This simple result for $Z=1$ is also found in the DCM, but the generation of the same probability and expected returns is not so evident, as it will be shown.

Table 1 lists some ICM results when only one single first prize is granted ($Z=1$). Given the single prize amount and the number of chips each player bears, the assigned money returns (in italics) are proportional to the chips counts. The same results are found for this case by the DCM procedure described below. Because of the chosen amount for the single first prize (100 arbitrary units), the proposed returns by ICM and DCM can be directly interpreted as percentages to win the prize (second row entry in italics).

### 3. A simple DCM procedure: the two player case

The origin of the DCM idea comes from a simple experiment involving two players that hold the same amount of chips. It is assumed that both players are equally skilled. So, two single outcomes are considered in an all-in scenario: the first player wins all the chips or the second one does it with the same probability. This is equivalent to a coin toss. It can be interpreted that the single outcome arises from a unique poker hand when two equally skilled players with the same stacks go all-in with the same probability to win, $1/2$. In both situations ties are not being considered as a possibility.

A variant of the previous experiment can be formulated when the amounts of chips are not the same and when two prizes are to be dealt, $P_1$ and $P_2$ (the second one is usually lesser than the first, and eventually it can be zero). This configuration induces to expand a game tree when considering all the possible outcomes after successive all-in situations. For instance, consider the tree of Figure 1. In the (leftmost) first node player 1 holds 1000 chips and player 2 only 500. As said above, two outcomes are possible. The first possibility is found when player 1 wins (first rectangular and upper terminal node in Figure 1) with probability $1/2$. The game ends here and player 1 collects all the chips and gets the first prize ($P_1$). Player 2 collects the second prize ($P_2$), if any. The other possibility arises if player 2 wins at the first game hand. In this case the amounts of chips are reversed (see Figure 1) and this occurs half of the time. At this spot two new possibilities are to be considered: if player 2 wins again (probability $1/2$ from the last node but cumulating a total probability of $1/2 \cdot 1/2 = 1/4$ from the beginning of the tree) the game ends and now is player 2 who receives the first prize. The other possibility has also a cumulated probability of $1/4$ and here player 1 recovers his/her original amount of chips. The tree expands from this non-terminal node but, because of the number of chips considered at the beginning, in this case the tree replicates itself in a recursive manner (Everett, 1957), and that is the meaning of the ellipsis in Figure 1.

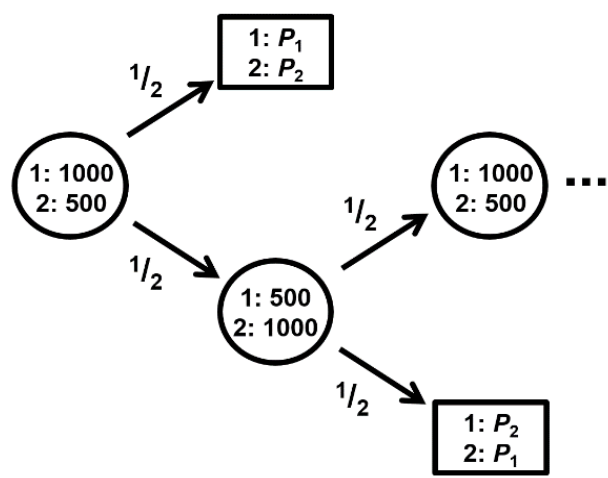

**Figure 1.** Game tree for two players involved in all-in situations. The first player initially has the double of chips than the second one. The ellipsis stands for a recursive repetition of the tree.

In this game, the probability $p$ of player 1 to be the final winner can be easily calculated. Following again the branches of the tree in Figure 1, this probability $p$ is $1/2$ (first terminal node) plus 0 (terminal node where player 2 wins the game) plus $1/2 \cdot 1/2$ of the probability attached to the repeated node. Note that the probability of this repeated node is again $p$, due to the recursive nature of the tree. That is:

$$p = \tfrac{1}{2} + \tfrac{1}{4}p.$$

This gives the DCM result of $p=2/3$, the same value obtained by the ICM approach calculated from the original chip amounts: $p=1000/(1000+500)$. As ties are not possible in this idealized game, the same result is obtained if at the reversed chips node (third node) it is considered that the probability for player 1 to win becomes $1-p$. Then, the equation becomes $p=1/2+1/2(1-p)$. A similar calculation gives the expected prize amount for player 1:

$$U_1 = \tfrac{1}{2}P_1 + \tfrac{1}{4}P_2 + \tfrac{1}{4}U_1,$$

where the term $1/4 P_2$ arises when player 1 loses at the lower terminal node of Figure 1. Hence, again the same result proposed by the ICM procedure is reproduced here:

$$U_1 = \tfrac{2}{3}P_1 + \tfrac{1}{3}P_2.$$

The game generates a more complex tree if the initial amounts of chips are in a distinct proportion than 2:1. In general, if the initial stacks are $a$ (player 1) and $b$ (player 2), the probability of player 1 to be the final winner can be depicted as the recursive tree of Figure 2,

$$p(a,b) = \tfrac{1}{2}p(a+e, b-e) + \tfrac{1}{2}p(a-e, b+e) \tag{1}$$

where $e=\min(a,b)$ is the effective stack.

Strictly, the unique termination conditions of the recursive formulation (1) are $p(0,y)=0$ and $p(x,0)=1$. Due to the equal playing skills attributed to each player, the condition $p(x,x)=1/2$ (virtual tie) can be also used to prune (correctly) some tree branches.

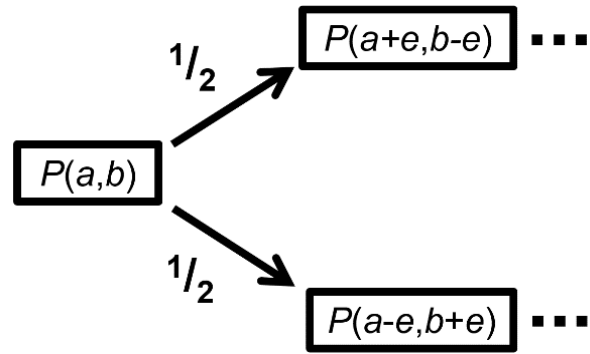

**Figure 2.** Recursive play tree involving two players with initial stacks *a* and *b*. The quantity *e*=min(*a*,*b*) is the hand confrontation effective stack.

During the generation of the tree, when assigning probability contributions to player 1, the multiplicative non-zero probability terms are being added path by path. This corresponds to consider only the paths having terminal nodes of the form *p*(*x*,0) or *p*(*x*,*x*). After the calculation, player 1 expected returns from the prizes are

$$U_1 = p(a,b)P_1 + \left[1 - p(a,b)\right]P_2. \qquad (2)$$

Note that, when recursively implementing formulas (1) and (2), the computational stack overflow error has to be avoided. For instance, if two players begin with stacks of 2 and 1 chips, an infinite path would be generated where the first player loses and wins alternatively (as found in Figure 1). In these cases it is enough to truncate the path at a certain deep (e.g. after a sequence of 50 hands or, equivalently, sequentially visited nodes). In general, the truncation terms are *p*(*x*,*y*)=1 if *x*>*y* (i.e., player 1 has more chances to win because of his bigger stack), *p*(*x*,*y*)=0 if *x*<*y* (player 1 has lesser chances to win) and, as said, $p(x,x)=\tfrac{1}{2}$. Of course, at any stage and according to the coincident values of DCM and ICM, for practical purposes it could be also considered the truncation term *p*(*x*,*y*)=*x*/(*x*+*y*), but previous options allow for a numerical demonstration of the ICM formula. Table 2 shows some numerical results obtained after the implementation of the recursive tree formulas (1) and

(2). The results are coincident with the ICM procedure because in the examples there are only two involved players.

| Initial Stacks | | Prizes | | Player 1 results | |
|---|---|---|---|---|---|
| Player 1 | Player 2 | 1rst | 2nd | Probability to win (%) | Expected prizes return |
| 1000 | 500 | 100 | 0 | 66.67 | 66.67 |
| 1000 | 500 | 100 | 50 | 66.67 | 83.33 |
| 1000 | 500 | 100 | 100 | 66.67 | 100 |
| 1000 | 100 | 100 | 50 | 90.91 | 95.45 |

**Table 2.** DCM results arising from procedures (1) and (2) involving two players with a distinct number of initial chips and one or two prizes. The results are coincident with those provided by the ICM procedure because here only two players are confronted.

## 4. General DCM procedure: the many players case

The DCM constitutes a simulation of a simplified and idealized Texas hold'em poker game. The general procedure consists into generate a recursive game tree. At each node *n* players enter a simulated confrontation where all of them go all-in and only one is the winner. The players are considered equally skilled, so any player can win and this forces the exploration of *n* possibilities. For each winner a new branch is created from the parent node and a new child node is formed. Hence, from a parent node involving *n* players, *n* new child branches arise leading to *n* child nodes, one for each winner. After each all-in confrontation takes place at the parent node and a player is declared to be the winner of the hand, the result is evaluated and the chips are redistributed accordingly in order to set up the child node. After each all-in result the players are partitioned into two groups: the one/s who still remain playing, i.e. still have chips and will participate in the next hand, and the one/s who get busted out and finish. The later one/s collect the respective prizes, if any, according to their final positions in the contest podium and, if necessary, resolving ties (see below). The former surviving players start a new all-in confrontation at the child tree nodes with the updated new stacks. The process continues expanding all the new nodes which, in turn, act as new parent nodes, until a termination node is reached (or the tree is pruned). Termination nodes are those where only a single player remains (and collects the first contest prize).

*4.1. The global DCM tree*

The game described in the previous paragraph generates the full or global game tree. It is assumed that *n* equally skilled players go all-in at each hand (and node). Of course, this is not realistic, but each *in silico* all-in situation can stand for a representation or simulation of a series of hands where all the players have entered a similar number of times and with equal mean probabilities to win. The probability for each player to win the hand is assumed to be 1/*n*, as they are equally skilled, and only a single player can be the winner. Here a main difference compared to the ICM is found: the probability to win a hand is not directly calculated from the number of chips (as said, ICM considers quotients of chip amounts). In the DCM, the probability to win is obtained by cumulating the probability terms being generated when advancing along the game tree branches. Of course, the number of chips each player bears influences the evolution of the game. This is so because, when a player wins a hand, this generates a redistribution of chips that promotes some players and harms others. Hence the adjective 'dependent' in the name of DCM.

The basic DCM procedure is depicted in Figure 3. The representation stands for a recursive DCM tree. The leftmost node shown can be either the first initial node or a representation of any intermediate one. In general, in a node *n* players have an initial positive number of chips in their respective stacks: $S_1$, $S_2$, ..., $S_n$. It is considered that all the *n* players go all-in and one and only one (one at a time) of them is the hand winner with probability 1/*n*. The *n* outcomes are considered (first player wins, or second player wins, ...) and this leads to the generation of *n* new tree branches leading to the corresponding child nodes.

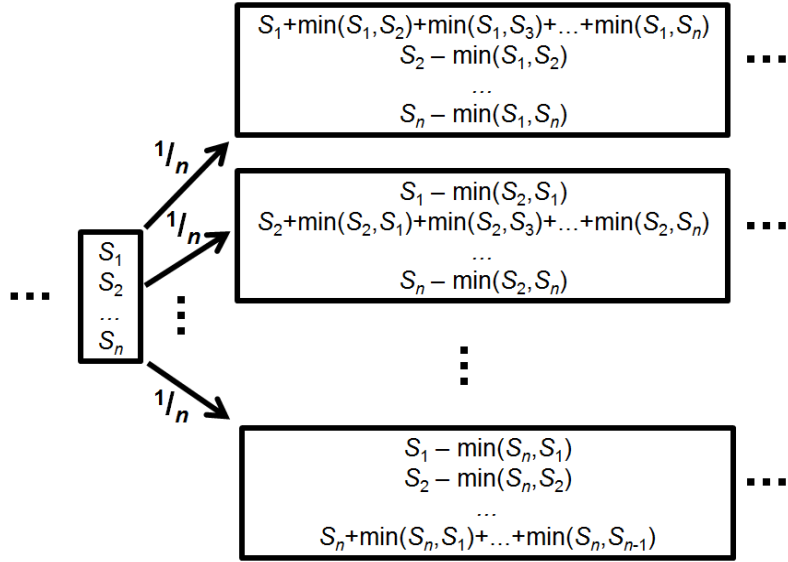

**Figure 3.** General tree expansion scheme of the generation of the recursive DCM game tree involving *n* players in a node with initial non-zero stacks $S_1, S_2, ..., S_n$. Each new child node and chips generated on the right corresponds to the event where one single player (one at a time at each node) wins the hand when all the participants went all-in in the parent node. At child nodes some players can reach bankruptcy after chips update. The total number of chips on the table is always kept constant.

Figure 3 shows how, at each child node, each winner $j$ ($j=1,n$) collects chips from all the other players, according to the respective effective stacks:

$$S_j \to S_j + \min(S_j, S_1) + ... + \min(S_j, S_{j-1}) + \min(S_j, S_{j+1}) + ... + \min(S_j, S_n).$$

Inside the same node, the other players $i \neq j$ lose chips:

$$S_{i \neq j} \to S_{i \neq j} - \min(S_i, S_j)$$

and in some cases (when $S_j \geq S_i$) the player $i$ gets bankrupted ($S_i$ becomes 0) and stops playing. At this stage the finishing players collect their prizes, if any, according to the reached position in the final podium (see below the treatment of bankruptcy and the resolution of the eventual ties that can occur). Afterwards, $n' \leq n$ players with the respective newly updated stacks still remain playing with a certain amount of chips, and a total of *n*' new child nodes are to be newly generated, one for each winner. Each one of the next child nodes are to be associated to a new multiplicative probability term equal to the new probability the remaining players have to win in the node, i.e., 1/*n*'.

The probability of the described successive events (hands) is the aforementioned 1/*n* term associated to each branch and multiplied by the previous terms coming from the previous path branches leading up to the original first node. The collected probability terms along a path (from the initial node when the initial *N* players start to play, up to a certain node) are of the form

$$p_k = \left(\tfrac{1}{N}\right)^p \ldots \left(\tfrac{1}{n}\right)^q \left(\tfrac{1}{n'}\right)^r \ldots \left(\tfrac{1}{n''}\right)^s, \quad N \geq n \geq n' \geq \ldots \geq n'', \qquad (3)$$

where *p*, *q*, *r*,..., *s* are non-null integers, i.e., $p_k$ is a product of powers of inverses of integer numbers (number of players surviving at each node). In the last term, *n*" is equal to the number of players confronted at the final path node (*n*"=1 if the path was not truncated and the branch was fully inspected). For every DCM tree, the sum of all the $p_k$ values at any stage always add 1. This sum accounts for the probability to find a winner. In the ideal case of an exhaustive tree search, were no truncations were forced, all the possible winners are finally identified.

Figure 4 stands for a global DCM game tree. It consists of an original parent node (leftmost one), the subsequent child nodes and, eventually, terminal tree nodes (drawn as squares) where only a single player remains collecting all the chips. This player is the winner of a particular path.

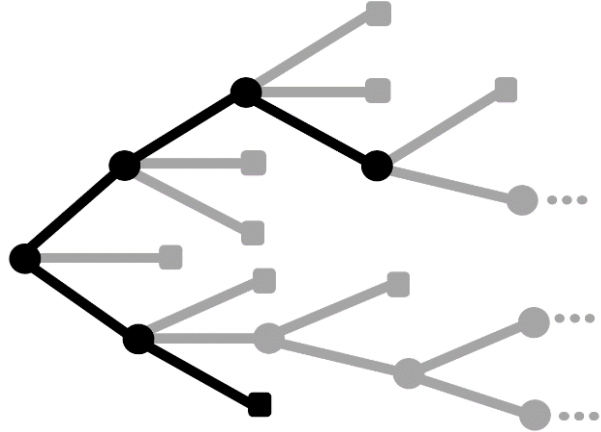

**Figure 4.** Example of a recursive DCM game tree involving several players. The lower black path and nodes stand for an example of a winning path. The winner is the player that survives at the squared terminal tree node. The upper black path ends with a play finalization node where some players may reach bankruptcy (one player in the depicted case), and others continue playing (at the child gray paths).

Normally, a DCM tree can be very deep (that is the meaning of the ellipsis) and it can be also infinite because there are branches where two or more players enter into a never ending sequence of alternative wins and losses. These branches involve paths with an infinite number of $1/n$ terms and their respective $p_k$ values are zero. In practical applications (e.g. when programming or generating the tree) these branches must be pruned at a certain stage in order to prevent computational stack overflows (and unaffordable code execution timings). Pruning is forced when the cumulated $p_k$ value is lesser than a pre-established threshold or when the reached tree node is deep enough, as explained below.

*4.2. Partial DCM game tree seen by a player*

Across the simulated confrontations, each player survives in distinct paths, sometimes reaching a final termination node or sometimes ending at intermediate nodes, when he/she founds bankruptcy. Hence, from a player

point of view there are two kinds of nodes and paths which are subparts of the global game tree:

a) The *terminal tree* nodes define players *winning paths* and are the nodes where one single player takes all the money at the final hand and becomes the winner of the game for this particular path (as an example, see the black lower path marked in Figure 4 and ending with a squared node).
b) The *finalization* nodes define players *losing paths* where the player finishes due to bankruptcy (see the black upper path marked in Figure 4 which ends with a rounded black node, the spot where a player loses its entire stack). In a finalization node the player can reach paid podium positions (e.g. if he/she has passed the bubble).

As said, each kind of path generates a probability term $p_k$. For example, the upper losing path identified in Figure 4 collects the probability term $1/3 \cdot 1/3 \cdot 1/3 = 1/27$. This is the probability to reach this spot. As an alternative, the lower winning path adds a probability winning term of $1/3 \cdot 1/3 = 1/9$ to a specific player. As said, the sum of all the winning contributions is 1. This sum can be partitioned according to assignments to the winning players. The fraction of cumulated winning probabilities gives the probability a player *i* has to win the contest:

$$PW_i = \sum_{k}^{\substack{\text{Winning paths} \\ \text{of player } i}} p_k,$$

and

$$1 = \sum_{i}^{\text{Players}} PW_i. \tag{4}$$

Each time a player reaches a winning or losing node a prize, $P_k$, is eventually assigned to him/her, according to the particular position achieved in the podium. Normally, early terminations will be assigned to a $P_k=0$ null prize. Conversely, the maximum prize is obtained at terminal nodes, at the end of a winning path.

The DCM prize contribution expectation for each player and path is the $p_k \cdot P_k$ product. In other words, each prize collected at a node is weighted by the corresponding node (path) probability term. Note that, in general, many of these terms will be zero (in certain paths the player does not reach prizewinning positions). The final DCM prize expectation for a player $i$, $DCM_i$, is obtained adding all the non-null $p_k \cdot P_k$ products:

$$DCM_i = \sum_k^{\text{Player } i \text{ finalization or termination nodes}} p_k P_k . \qquad (5)$$

The sum of all the DCM player expectations gives the total amount of the contest pool prizes, $C_k$:

$$\sum_i^{\text{Players}} DCM_i = \sum_k^{\text{Prizes}} C_k .$$

The nominal contest prizes, $C_k$, can or cannot coincide with assigned prizes along the tree, $P_k$. In a contest where $N$ initial players are participating, only $Z \leq N$ assigned monetary prizes (podium positions) are specified: $C_1 \leq C_2 \leq ... \leq C_Z$. Note the particular sorting convention used here: the bigger the prize sub index the better the podium position and monetary reward. For practical computational purposes $N$ values ($P_1$, $P_2$, ..., $P_N$) are to be defined as available prizes. If $N>Z$ this is easily done taking the $C_k$ ones and extending with zeroes the list of $P_k$ values, if necessary:

$$\{P_k\}_{k=1,N} = \{0_k\}_{k=1,N-Z} \cup \{C_k\}_{k=1,Z} = \{0_1, 0_2, ..., 0_{N-Z}, C_1, C_2, ..., C_Z\}, \quad \text{for } N>Z$$

and

$$\{P_k\}_{k=1,N} = \{C_k\}_{k=1,Z}, \quad \text{for } N=Z.$$

There are situations where the assigned prize to a player in a node does not coincide with any of the pre-established $P_k$ values. This is due to the ties that can be generated when several players reach bankruptcy (see below). This situation forces one or more prizes to be split in certain nodes. In any case, the first prize is always assigned at a terminal winning node. Hence, this prize is never split except for the cases where a branch is pruned.

The ICM procedure does not deal explicitly with tie situations, but it does implicitly. Within the ICM context, when doing calculations when generating all the possible podium variations, ties are solved by means of assignations of additive equal terms to distinct players.

*4.3. DCM treatment of bankruptcy and eventual ties in play finalization nodes*

When a particular player wins the hand, some other players can get into bankruptcy at the same time, and the eventual prizes are to be distributed correctly in the node and treating ties, if necessary. The DCM definition presented here does not rely on an expeditious or simple procedure to resolve bankruptcy, but on a common, natural and fair one based upon podium positioning and administrating possible ties. The factors determining the prizes distribution after bankruptcy are the original stacks the players had when entering the node, i.e., just before the assignation of the winning player. Let us assume that, in a node, $L \geq 1$ players lose all their chips against a certain player and that they entered previously into the node with (sorted) stacks $S_1 \leq S_2 \leq ,..., \leq S_L$. The stacks ordering establishes the sequence of prize assignations. A total of $L$ sorted prizes are to be shared and inside this node are re-numbered correlatively: $P_1 \leq P_2 \leq ,..., \leq P_L$. Note that, previously, maybe other minor prizes ($\leq P_1$) were already assigned (and exhausted) in other previous nodes of the same path, and that some prizes are void if they are beyond the number of paid contest podium positions. The assignation of prizes is done by forming groups or lots of equal-sized stacks. The first group consists of $1 \leq T_1 \leq L$ tied (equal) stack players, and the second involves $1 \leq T_2 \leq L$ ties, and so on up to the last group of $1 \leq T_M \leq L$ ties. All these groups are complementary and satisfy the condition $T_1 + T_2 + ... + T_M = L$. Note that each group have, at least, one player.

All the players of the same group have to receive the same amount of money. Each player of the first group of tied people receives the amount

$$U_{1:T_1} = \frac{P_1 + P_2 + \ldots + P_{T_1}}{T_1},$$

i.e., the sum of the first series of $T_1$ prizes divided by the number of tied players of the group, as all those players will share (in the actual particular path) the same final contest podium position. Similarly, each player of the second tied group receives the amount

$$U_{T_1+1:T_1+T_2} = \frac{P_{T_1+1} + P_{T_1+2} + \ldots + P_{T_1+T_2}}{T_2}.$$

The groups of tied players in a node are paid successively, eventually accessing to higher prizes up to the final group. Each player of the last group will receive the amount

$$U_{L-T_M+1:L} = \frac{P_{L-T_M+1} + P_{L-T_M+2} + \ldots + P_{L-1} + P_L}{T_M}.$$

The simple Algorithm 1 listed at the end of the Appendix shows how the ties are solved in a node when $L$ players get into bankruptcy simultaneously in a node. The procedure here simply defines the respective player returns, $\{U_i\}_{i=1,L}$. Note that the terms $U_i$ coincide with the $P_i$ ones if $T_i=1$. When two or more players tie ($T_i>1$), the $P_i$ terms are split and equation (5) formally becomes the following general one:

$$DCM_i = \sum_k^{\text{Player } i \text{ finalization or termination nodes}} p_k U_k.$$

## 5. DCM example results

Algorithms implementing the complete DCM procedure are discussed in the Appendix. Table 3 shows several example results. If two or more prizes are to be negotiated, DCM and ICM assignations do not coincide when more of two players participate in the tournament. As it can be checked in Table 3, the percentage of difference respect to the ICM value of the first player is always positive. This percentage can be lesser than the absolute values of the negative percentages assigned to the players with small stacks, but it has to be taken into account that the first percentage is referenced to a higher monetary value.

| Players & Prizes | Player 1 | Player 2 | Player 3 | Player 4 | Player 5 |
|---|---|---|---|---|---|
|  | 1000 | 500 | 100 |  |  |
| 3 | 78.79 | 58.33 | 12.88 |  |  |
| 100 / 50 / 0 | *80.79* | *60.67* | *8.54* |  |  |
|  | +2.5% | +4% | -33.7% |  |  |
|  | 1000 | 500 | 100 |  |  |
| 3 | 79.28 | 59.79 | 20.93 |  |  |
| 100 / 50 / 10 | *80.88* | *61.66* | *17.46* |  |  |
|  | +2.0% | +3.1% | -16.6% |  |  |
|  | 1000 | 800 | 500 | 100 |  |
| 4 | 58.47 | 50.35 | 33.94 | 7.24 |  |
| 100 / 50 / 0 / 0 | *65.28* | *53.54* | *25.98* | *5.20* |  |
|  | +11.6% | +6.3% | -23.4% | -28.3% |  |
|  | 1000 | 800 | 500 | 200 | 100 |
| 5 | 53.96 | 45.95 | 30.70 | 12.88 | 6.52 |
| 100 / 50 / 0 / 0 / 0 | *61.08* | *47.59* | *24.19* | *12.58* | *4.57* |
|  | +13.2% | +3.6% | -21.2% | -2.3% | -29.9% |
|  | 1000 | 800 | 500 | 200 | 100 |
| 5 | 56.02 | 48.39 | 33.67 | 14.52 | 7.40 |
| 100 / 50 / 10 / 0 / 0 | *62.59* | *50.65* | *28.26* | *13.08* | *5.42* |
|  | +11.7% | +4.7% | -16.1% | -9.9% | -26.8% |

**Table 3.** Example results obtained with ICM and DCM procedures in prize negotiations. The entries in the first column are the number of players and the offered podium prizes. On the right, each row contains the initial players' chip amounts, the ICM proposal, the DCM alternative (in italics) and the percentage of difference respect to the classical ICM value.

All in all, the ICM constitutes a sort of shrinkage of DCM values. There is a transfer of the classical ICM money proposal from players with small stacks to the players having bigger stacks. This is conforming to a usual desire: sometimes the players with small stacks tend to ask for ICM negotiations. This may be due to the fact that they 'know' that ICM will assign more money than the deserved or fair one. In other words, ICM is benevolent with the small stacked players. The author of this article wonders about the heuristic feeling of players in relation to the fair or unfair level of the DCM proposals. Will the players be conformable with DCM assignations? Will the players admit as objective the DCM results? Possibly they should, as the aim of the definition of DCM is to provide with a fair-by-construction tool. As far as the knowledge of this author goes, DCM procedure is the unique protocol that ensures the players are treated having identical playing skills. The final differences among monetary assignations are only due to the initial unbalanced stacks. That is the reason to use the adjective 'dependent' in DCM name.

| Prized podium position | Player assignations from stacks | | |
|:---:|:---:|:---:|:---:|
| | **Stack of 1000 chips** | **Stack of 500 chips** | **Stack of 100 chips** |
| 1 | *62.50* | *21.25* | *6.25* |
| 2 | 32.58 | 54.17 | 13.26 |
| | *36.58* | *58.83* | *4.58* |
| 3 | 4.92 | 14.58 | 80.49 |
| | *0.92* | *9.92* | *89.17* |

**Table 4.** ICM (upper value in each row) and DCM (lower value, in italics) results obtained for 3 players when disputing for a single prize. This unique prize is of 100 units and is assigned to a particular single podium position (first, second or third) despite all the players disregards this and always play to reach the first position. In this way the DCM and ICM vertical partitions can be read as percentage probabilities the player has to finish in the specified podium position. For the first winning podium position the DCM and ICM results coincide, as said in text.

Table 4 presents DCM and ICM prizes distribution according to a special (and artificial) prizes structure. At each row there is only one single prized podium position of 100 units. Hence, the DCM and ICM values are again interpreted here as probabilities to receive the prize or, equivalently, to win the contest. The players have initially 1000, 500 and 100 chips (the same as in the first example of Table 3), respectively, and always play to reach the highest podium position (as usual!). The first row (winning podium position 1) shows the common case: the single prize is assigned to the first podium place. In the other two rows, only the second (or only the third) podium positioned player receives the single prize of 100 units. Of course this is an unrealistic and counterintuitive scenario, but the calculations (which have been forced slightly manipulating the computer codes) show an interesting feature: the numbers in each column are the probabilities the player has to reach one of the possible podium positions (assuming the player disregards the particular prize structure and always plays to reach the first position). In the first row the DCM and ICM results coincide (and are proportional to the chip amounts), as said above. The results of DCM (in italics) and ICM differ in the other two rows. Horizontal sums for each method add 100, and also do the vertical sums for each method. The vertical reading along each column indicates the probability the player has to end in first, second or third position according to each method. As it can be seen, DCM assigns to the short stacked player more probability to end at the third position.

The differences in probabilities assignation to final podium positions (Table 4) conditions the final differences between assigned DCM and ICM monetary values (Table 3). The proposed prize distributions of Table 3 can be obtained from adequate additions of probabilities multiplied by prize assignations. The first row entries of Table 3 can be reproduced taking into consideration the probabilities of Table 4 and weighting them with the assigned prizes to each position in Table 3 (100 units for first, 50 for second and 0 for the third player). For instance, the DCM monetary assignation for the first podium position (80.79 in Table 3) is obtained considering the probability of being first (62.50% in Table 4) times the first prize of 100, plus the probability of being second (36.58%) times the second prize of 50 units, plus the probability of being third (0.92%) times the third prize of 0 units. The complete linear combination is

$$DCM_{1st} = 0.6250 \times 100 + 0.3658 \times 50 + 0.0092 \times 0 = 80.79.$$

Equivalent calculations can be performed for the other podium positions and players, and the procedure also applies for the ICM procedure.

The new concept of DCM is interesting from poker players training (as in the case of ICM training programs) or decision tacking purposes. The concept ensures that the player receives an amount totally compatible with his/her stack and assuming perfect equality and playing skills relative to other players. Viewing DCM as a fair tool, if a player knows about differences among skills relative to other players, from the DCM approach he/she can heuristically extrapolate or anticipate (possible) future results. Accepting this idea leads to admit that the players are not treated by ICM as being equally skilled. For the cases where ICM and DCM differ in assignations, the first method conditions its result to the chip amounts. In fact, within the ICM framework, the players with smaller stacks are treated as having a little advantage relative to the other players. This originates the differences in monetary distributions when comparing results with the DCM method.

**6. ICM and DCM differences trigger distinct tournament strategies!**

The suggested actions by ICM and DCM can enter in conflict. This is not only a matter of heuristic feeling about the correctness or nature of both methods. Regarding the calculation of monetary expected values, the use of one method or another in order to set strategies in tournaments can lead to very distinct actions. This will be exposed here with an example.

Let us assume four players remain in a final poker table: Villain at the SB, Hero at the BB, and two others (CO, BTN). The initial stack sizes are 1000, 1000, 2000 and 3000, respectively. Hero put the $200 big blind and CO and BTN folded. Then, the Villain at the SB proposes an all-in pushing 1000 chips. The

prize pool if of $100 and pays $50 for first, $30 for second and $20 for third classified. Should the Hero call or should the Hero fold?

The Hero is in trouble because he/she knows that the Villain usually proposes all-in when this person holds good pocket cards. The Hero estimates that his/her equity against the Villain's range is about of only 40%. Does the intuition of the reader point to fold? Numbers will say.

According to ICM, Hero's monetary equity of a call/win is $31.86 (stacks of 0, 2000, 2000 and 3000) and, of course, it falls to $0 in case of a call/lose. The weighted value for the call considering the chances the Hero has to win the hand is $12.74. On the other side, if Hero folds, the stacks will be of 1200, 800, 2000 and 3000 chips, and the Hero has now an ICM monetary expectation of $15.23. According to the ICM calculations, as $12.74<$15.23, Hero should fold in order to keep a maximal long run expectation.

The calculations with DCM differ qualitatively in this selected example. Hero's DCM monetary equity of a call/win is $30.71 and $0 for a call/lose. The equity-weighted value is $12.29, a similar value to the ICM one. But if Hero folds the DCM monetary expectation is only of $9.57. Hence, according to the DCM calculations, as $9.57<$12.29, Hero should accept the all-in.

In this example, the important difference between ICM and DCM is due mainly to the effect DCM has in short stacks. Both methods, ICM and DCM, recommend a fold for equities ≤31% and both recommend accepting the all-in for equities ≥48%. In the intermediate range only DCM recommends going all-in. Qualitatively speaking, the DCM message in this example is simple: player with short stacks are shorter than they think if consulting ICM... and they must accept or go all-in more frequently than they usually do.

The author of this work wonders if DCM will provoke changes in (professional) players' strategies. Can DCM become a new paradigm for post-game analysis? It should if players agree that DCM is based in a more realistic calculation procedure than ICM.

All the commented DCM computer codes in this text and executables are available upon request to the author. Information will be updated at iqcc.udg.edu/~emili/Poker/DCM/

## 7. Conclusions

The Dependent Chip Model (DCM) has been defined in the context of Texas hold'em poker and constitutes a more realistic alternative to the Independent Chip Model (ICM). The procedure ensures to the participants a fair result only depending on the amount of chips. This is so because the players are treated as being always equally skilled during a play simulation. When only a single first prize is to be assigned both procedures, ICM and DCM, return the same monetary assignations. The same occurs when only two players are confronted. In the other cases (more than two players, more than two remunerated podium positions, single prizes artificially assigned to non-first podium positions) the procedures propose distinct assignations. DCM tends to assign more money to the first podium positions, and less to the final ones in comparison to ICM procedure. Due to the DCM design, the obtained results can be interpreted as a demonstration that ICM gives some skill advantage to short stacked players. A simple prize decomposition procedure has been also presented which allows the calculation of the probabilities each player has to end at every podium position. In turn, this lead to obtain the ICM and DCM assignations in an additive manner. Finally, an example has been presented, showing how ICM and DCM can recommend distinct actions to tournament players. This opens a door for further investigations.

## 8. References

Diaconis, P.; Ethier, S. N. (2020) Gambler's Ruin and the ICM. https://arxiv.org/abs/2011.07610.

Everett, H. (1957) Recursive games. In Contributions to the Theory of Games Vol. III, Vol 39 of Annals of Mathematical Studies.

Ganzfried, S. and Sandholm, T. (2008) Computing an Approximate Jam/Fold Equilibrium for 3-player No-Limit Texas Hold'em Tournaments. Proc. of 7th Int. Conf. on Autonomous Agents and Multiagent Systems (AAMAS 2008), Padgham, Parkes,Müller and Parsons (eds.), May, 12-16., 2008, Estoril, Portugal.

Gilbert, G. T. (2009) The independent chip model and risk aversion. https://arxiv.org/abs/0911.3100.

Miltersen, P. B.; Sørensen, T. B. (2007) A near-optimal strategy for a heads-up no-limit Texas Hold'em poker tournament. In International Conference on Autonomous Agents and Multi-Agent Systems (AAMAS). DOI 10.1145/1329125.1329357.

Sklansky, D. (2002) Tournament Poker for Advanced Players. Two Plus Two Publishing.

**Appendix**

The codes presented here implement all the parts of DCM procedure: the tree generation and tracking, bankruptcy (and eventual ties) treatment, and update of $PW_i$ (probability a player has to win the contest) and $DCM_i$ (DCM contributions) variables for each player. The algorithms are not optimized nor they present check or control sentences. This is so in order to maintain a clear exposition of the concepts and ideas beyond DCM, linked to the main text. More information and codes are posted at iqcc.udg.edu/~emili/Poker/DCM/.

The routines use a track of players' identities, $\{w_i\}$. This is necessary because some players will eventually be sorted (according to stacks) or removed from the recursive search when traveling across the game tree. The identity variable management ensures that the variables $PW_i$ and $DCM_i$ always are updated properly for any player. The initial stacks are assigned to particular player identities (among the total of *N* players). Hence, the $w_i$ values are all distinct and take values between 1 and *N*.

Algorithm 1 is generalized and implemented in terms of a routine, Solve_Bankruptcy() which is codified by the Algorithm 2 below. This routine takes into account the progressive update of the DCM values for each one of the *L* players that get into bankruptcy. The total number of players is *N* at the beginning of the contest, the number of still active player in a node is *n* and a subset of *L* of them are the ones that lose all their chips. The current $DCM_i$ (*i*=1,*N*) values of all the players are passed to the routine, but only those of the *L* active players that get into bankruptcy are updated (the ones identified by the transferred $w_i$ terms). The update takes into account the same monetary attribution $U_i$ found in Algorithm 1 but weighting it by the probability *p* to reach the node. The processes of sorting stacks and prizes are to be done avoiding collateral effects to outer codes (that is the reason why new copy variables are defined). When sorting stacks, always the respective player identities are also to be sorted accordingly.

The routine Solve_Bankruptcy() is employed by Algorithm 3 below, which implements the DCM procedure. The routine New_Hand() is recursive and generates the tree expansion by means of a deep first tree search. The input variables are the number of active players in a node (*n*), their identities $\{w_i\}_{i=1,n}$, the set of respective stacks $\{S_i\}_{i=1,n}$, and the prizes *still* to be assigned $\{P_i\}_{i=1,n}$. The variables $PW_i$ (player winning probabilities) and $DCM_i$ (DCM values) always refer to the complete set of *N* players that entered the first node of the tree. These two accumulators are being updated during the game tree tracking. To be properly actualized, these variables need the use of the players' identities.

The code of Algorithm 3 needs previously two parameters to be defined: *MD*, the maximal acceptable tree deep search and *ProbMin*, the minimal acceptable probability for a node. In this work the considered values were *MD*=50 and *ProbMin*=$10^{-15}$ that serve to force the tree search truncation. When the limit condition is met (see code), a bankruptcy situation for all the still active players in the node is forced. This ensures at this stage to share the remaining prizes according to the actual stacks, as the game tree will not be further explored from this branch.

If the prizes are sorted from the beginning by Algorithm 3, then Algorithm 2 does not need to sort them again when called. Note that, in order to avoid collateral effects, in Algorithm 3 the call to the routine Solve_Bankruptcy() is performed defining new variables for player stacks ($S''_i$) and proper identities ($w''_i$). The assignation of the first prize to the final winner at a path ending node (when $n=1$) is performed using a single sentence ($DCM_{w_1} = DCM_{w_1} + pP_1$). This corresponds to the particular case of calling again Algorithm 2 but involving only a single player ($L=1$, the winner with identity $w_1$) and a single remaining prize ($P_1$). The Algorithm 3 also manages the probability terms appearing in formula (3). These probabilities are carried in $p$ and $p'$ variables. In this code the cumulated winning probabilities of each player is obtained ($PW_{w_1} = PW_{w_1} + p$). Despite the update of $PW_i$ terms is not compulsive in order to get the DCM ones, it is worth nothing that the actual sum of $PW$ values signals the degree or fraction of completion of the DCM calculation. This is so because, according to equation (4), at the end of the full tracking of the playing tree the sum must add 1. By using the proposed codes, this value could not be reached if too early or too much pruning actions are done.

Algorithms:

Set *L*, the number of players that get into bankruptcy in this node.
Consider the sorted stacks, $\{S_i\}_{i=1,L}$ ($S_i \leq S_j$, $\forall i<j$.), they had when entering the node.
Consider the prizes to be shared in this node: $\{P_i\}_{i=1,L}$.

Sort prizes from lower to higher: $P_i \leq P_j$, $\forall i<j$.
*i*=0
*PL*=1. Actual prize level. This will increase up to *L*.
Do while *i*<*L* (Ascending value of *i* from 1 up to *L*).
    *i*=*i*+1
    If *i*<*L* then
        *k*=*i*+1
        Do
            If $S_i=S_k$ then (tied players)
                If *k*=*L* Exit do
                *k*=*k*+1
            else
                *k*=*k*-1
                Exit do
            End if
        End do
    else
        *k*=*i*
    End if

$$U(i:k) = \frac{\sum_{j=PL}^{PL+k-i} P_j}{k-i+1}$$

    *PL*=*PL*+*k*-*i*+1
    *i*=*k*
End do

**Algorithm 1.** Implementation of the money share ($U_1$, $U_2$,... $U_L$) when *L* players get bankrupted in a node they entered with stacks $S_i$ (it is assumed that the stacks are sorted in advance). A total of *L* prizes are distributed. Each block or range *i*→*k* embraces a single player or a group of tied players (i.e. with equal initial stacks) that will receive the same amount of money.

Partial previous updated DCM values, $\{DCM_i\}_{i=1,N}$, are known for all the $N$ players.
Set $p$, the probability attached to the node.
Set $L$, the number of players that get into bankruptcy in this node.
Consider their entering stacks: $\{S_i\}_{i=1,L}$ and the corresponding player identities $\{w_i\}_{i=1,L}$.
Consider the prizes to be shared (some can be zero): $\{P_i\}_{i=1,L}$.
Call Solve_Bankruptcy ($p$, $N$, $\{DCM_i\}_{i=1,N}$, $L$, $\{w_i\}_{i=1,L}$, $\{S_i\}_{i=1,L}$, $\{P_i\}_{i=1,L}$)

Routine Solve_Bankruptcy($p$, $N$, $\{DCM_i\}_{i=1,N}$, $L$, $\{w_i\}_{i=1,L}$, $\{S_i\}_{i=1,L}$, $\{P_i\}_{i=1,L}$)
    Sort stacks from lower to higher: $S_i \le S_j$, $\forall i<j$. Doing that, drag the players' identities.
    Sort prizes from lower to higher: $P_i \le P_j$, $\forall i<j$.
    $i=0$
    $PL=1$
    Do while $i<L$
        $i=i+1$
        If $i<L$ then
            $k=i+1$
            Do
                If $S_i=S_k$ then
                    If $k=L$ Exit do
                    $k=k+1$
                else
                    $k=k-1$
                    Exit do
                End if
            End do
        else
            $k=i$
        End if
        For $m=i,k$

$$DCM_{w_m} \leftarrow DCM_{w_m} + p \frac{\sum_{j=PL}^{PL+k-i} P_j}{k-i+1}.$$ Update the DCM values of involved players.

        End for
        $PL=PL+k-i+1$
        $i=k$
    End do
End Solve_Bankruptcy

**Algorithm 2.** Algorithm implementing the update of DCM money share when $L$ players get to bankruptcy in a node. A total of $L$ prizes are to be distributed. Groups of tied players (i.e. with equal stacks) are embraced by the blocks of ranges $i \rightarrow k$ (each range of players receive the same amount of money weighted by the probability $p$ to reach the node). The considered stacks are those of the players had before the hand showdown that occurred in the node.

Parameter: Set *MD*, the maximal acceptable tree deep search.
Parameter: Set *ProbMin*, the minimal acceptable probability for a node.
Total initial number of players: *N*.
Identities initialization: $w_i=i$, $\forall i=1,N$.
The DCM values: $DCM_i=0$, $\forall i=1,N$.
Players' winning probabilities: $PW_i=0$, $\forall i=1,N$.
$p=1/N$. Next node probability.
Set the players' stacks: $S_i>0$, $\forall i=1,N$.
Set the prizes for the *N* podium positions: $P_i$, $\forall i=1,N$, sorted from lower to higher.
*Deep*=1. Actual tree deep
*n*=*N*, at the first node, the set of *n* involved players in the hand are all the *N* players.

recursive routine New_Hand(*Deep*, *p*, *n*, $\{w_i\}_{i=1,n}$, $\{S_i\}_{i=1,n}$, $\{P_i\}_{i=1,n}$, *N*, $\{PW_i\}_{i=1,N}$, $\{DCM_i\}_{i=1,N}$)
    If *Deep*=*MD* or *p*<*PromMin* then (Pruning: tree search termination is forced)
        Call Solve_Bankruptcy (*p*, *N*, $\{DCM_i\}_{i=1,N}$, *n*, $\{w_i\}_{i=1,n}$, $\{S_i\}_{i=1,n}$, $\{P_i\}_{i=1,n}$)
        Return
    End if
    If *n*=1 then (Termination node. Always one single winner)
        Assign first prize to winner: $DCM_{w_1} = DCM_{w_1} + pP_1$.

        Update the probability to win for this player: $PW_{w_1} = PW_{w_1} + p$.

        Return
    End if
    For *j*=1 to *n* (The winning player is the *j*-th one)
        Initialize new players' stacks as $S'_i=S_i$, $\forall i=1,n$.
        All players go all-in and declare player *j* as unique winner.
        Re-arrange stacks accordingly (Figure 3): new stacks are $\{S'_i\}_{i=1,n}$.

        If *L*>0 players get to bankruptcy (i.e., they have $S'_i=0$) then
            Identify the *L* players:
                -Set their identities ($w''_i$).
                -Set their original stacks ($S''_i$) when entering the node.
            Set the prizes ($P_i$) to distribute.
            Call Solve_Bankruptcy(*p*, *N*, $\{DCM_i\}_{i=1,N}$, *L*, $\{w''_i\}_{i=1,L}$, $\{S''_i\}_{i=1,L}$, $\{P_i\}_{i=1,L}$)
        End if

        Set number of remaining players (*n'*) with stacks $S'_i>0$ and identities $w'_i$.
        Set the remaining prizes $P'_i$.
        *p'*=*p*/*n'*: Next node probability term.
        *Deep'*=*Deep*+1
        Call New_Hand(*Deep'*, *p'*, *n'*, $\{w'_i\}_{i=1,n'}$, $\{S'_i\}_{i=1,n'}$, $\{P'_i\}_{i=1,n'}$, *N*, $\{PW_i\}_{i=1,N}$, $\{DCM_i\}_{i=1,N}$)

    End for
End New_Hand
The amounts $DCM_i$ are the DCM player assignments.
The $PW_i$ terms indicate the probability each player has to be the absolute winner.

**Algorithm 3.** Implementation of the DCM procedure.

___________________